\documentclass{osa-article}

\journal{oe}
\usepackage{lineno}
\begin{document}

\title{Optical force and torque on small particles induced by polarization singularities}

\author{Jie Peng,\authormark{1} Shiqi Jia,\authormark{1} Chengzhi Zhang\authormark{1}, and Shubo Wang\authormark{1,2,*}}

\address{\authormark{1}Department of Physics, City University of Hong Kong, Tat Chee Avenue, Kowloon, Hong Kong, China\\
\authormark{2}City University of Hong Kong Shenzhen Research Institute, Shenzhen, Guangdong 518057, China\\}

\email{\authormark{*}shubwang@cityu.edu.hk} %% email address is required



\begin{abstract}
Optical forces in the near fields have important applications in on-chip optical manipulations of small particles and molecules. Here, we report a study of optical force and torque on small particles induced by the optical polarization singularities of a gold cylinder. We show that the scattering of the cylinder generates both electric and magnetic C lines (i.e., lines of polarization singularities) in the near fields, and the C lines can induce complex force and torque on a dielectric/magnetic particle. The force and torque manifest dramatic spatial variations with interesting symmetry properties, providing rich degrees of freedom for near-field optical manipulations. The study, for the first time to our knowledge, uncovers the effect of optical polarization singularities on light-induced force and torque on small particles. The results contribute to the understanding of chiral light-matter interactions and can find applications in on-chip optical manipulations and optical sensing.
\end{abstract}

%%%%%%%%%%%%%%%%%%%%%%%%%%  body  %%%%%%%%%%%%%%%%%%%%%%%%%%
\section{Introduction}
Light carries both linear momentum and angular momentum, and it can induce forces and torques on objects due to momentum conservation. Optical forces and torques can enable precise manipulations of small particles in the micron or nano scale with rich applications in physics, chemistry, and biology \cite{berns_use_1989,chaumet_optical_2002,la_porta_optical_2004,ren_wavelet_2017,bradac_nanoscale_2018}. Recently, optical forces induced by the near fields of optical structures have attracted growing interests due to their large magnitudes and rich degrees of freedom deriving from the vector properties of the fields \cite{canaguier2014transverse,bliokh_extraordinary_2014,wang_lateral_2014,rodriguez-fortuno_lateral_2015,petrov_surface_2016,wang_strong_2016,lu_light-induced_2017,ivinskaya_optomechanical_2018}, which have important applications in on-chip optical manipulations of molecules, cells, and viruses \cite{hayat_lateral_2015}. In particular, near-field chiral light-matter interactions can give rise to intriguing lateral optical forces that can be employed to separate chiral particles \cite{wang_lateral_2014,rodriguez-fortuno_lateral_2015,hayat_lateral_2015,canaguier2015plasmonic,kalhor2016universal,pendharker2018spin,shi_chirality-assisted_2020,shi_optical_2020}. Therefore, exploring the optical forces induced by chiral near fields (i.e., circularly polarized fields) is of critical importance to the development of novel manipulation functionalities. 

The chiral fields in optical structures are closely related to the C points, corresponding to spatial locations at which the field is circularly polarized and the orientation of the major axis of polarization ellipse is ill-defined  \cite{noauthor_lines_1983,noauthor_wave_1987}.  The C points can emerge in various optical systems including light scattering by simple structures  \cite{garcia-etxarri_optical_2017,yu_kuznetsov_three-dimensional_2020,chen_extremize_2021,peng_polarization_2021,liu_topological_2021,peng_topological_nodate}, and they have been observed experimentally  \cite{noauthor_observations_1990,egorov_experimental_2008,burresi_observation_2009,bauer_observation_2015,doeleman_experimental_2018,zhang_observation_2018,chen_observing_2019,lustig_identifying_2020}. The C points are topological defects of the optical polarization fields with interesting topological properties characterized by quantized integer/half-integer indices (i.e., winding numbers)\cite{liu_topological_2021}. Remarkably, the topological properties of near-field C points have subtle connections with the topology of optical structures\cite{peng_topological_nodate}. Moreover, it has been shown that C points have subtle relations with bounded states in the continuum, exceptional points in the non-Hermitian systems \cite{li_geometric_2019,chen_evolution_2020}, and geometric phases \cite{bliokh_geometric_2019,maji_geometric_nodate}. In the three-dimensional (3D) space, the C points usually form lines, and these C lines have complex configurations giving rise to structured chiral near fields with applications in chiral light-matter interactions and chiral sensing\cite{jia_chiral_nodate}. 

In this work, we theoretically investigate the optical force and torque induced by C lines on subwavelength dielectric/magnetic particles. We consider the C lines that emerge in both electric and magnetic near fields of a gold cylinder excited by a linearly polarized plane wave. We show that the induced optical force and torque have dramatic variations along the C lines, enabling nontrivial manipulations of the small particles. By applying dipole approximations, we find that the force is mainly dominated by the gradient force deriving from the strong variation of the fields around the C lines, while the torque is directly proportional to the spin densities on the C lines. In addition, we uncover interesting symmetry properties of the force and torque fields.

The paper is organized as follows. In Sec.\ \ref{Sec. 2}, we discuss the properties of the C lines in the electric and magnetic near fields of the gold cylinder. In Sec.\ \ref{Sec.3}, we present numerical and analytical results for the optical force and torque on small dielectric/magnetic particles locating on the two types of C lines. We also discuss the relationships between the force/torque and the properties of the C lines. We then conclude in Sec.\ \ref{Sec.4}.

\section{\label{Sec. 2}Polarization singularities in the near fields of a gold cylinder}
A general 3D electric field can be expressed as $\mathbf{E}=(\mathbf{A}+\mathrm{i}\mathbf{B}) e^{\text{i} \theta}$, where $\theta=\arg (\mathbf{E} \cdot \mathbf{E}) / 2$ is a proper phase, $\mathbf{A}= \operatorname{Re}\left[\mathbf{E} \sqrt{\mathbf{E}^{*} \cdot \mathbf{E}^{*}}\right]/|\sqrt{\mathbf{E} \cdot \mathbf{E}}|$ and $ \mathbf{B}=\operatorname{Im}\left[\mathbf{E} \sqrt{\mathbf{E}^{*} \cdot \mathbf{E}^{*}}\right]/|\sqrt{\mathbf{E} \cdot \mathbf{E}}|$ are the major and minor axes of the polarization ellipse \cite{berry_index_2004}. The C points of the electric field correspond to the spatial points at which $\mathbf{A}=\mathbf{B} \neq \mathbf{0}$. The C points of the magnetic field can be defined in a similar way. These points are topological singularities of the polarization fields, and they can be characterized by a topological index defined as $I=\frac{1}{4 \pi} \oint d \phi$, where $\phi$ is the azimuthal angle on the Poincaré sphere for the in-plane polarization (in the plane of the polarization ellipse at the C points), and the integral is evaluated on a small loop enclosing the C points. The C points form C lines in 3D space, and the topological index of the C lines is generally $I=\pm 1 / 2$ without the protection of any symmetry  \cite{berry_index_2004}.

\begin{figure}[t!]
\centering\includegraphics[width=1\linewidth]{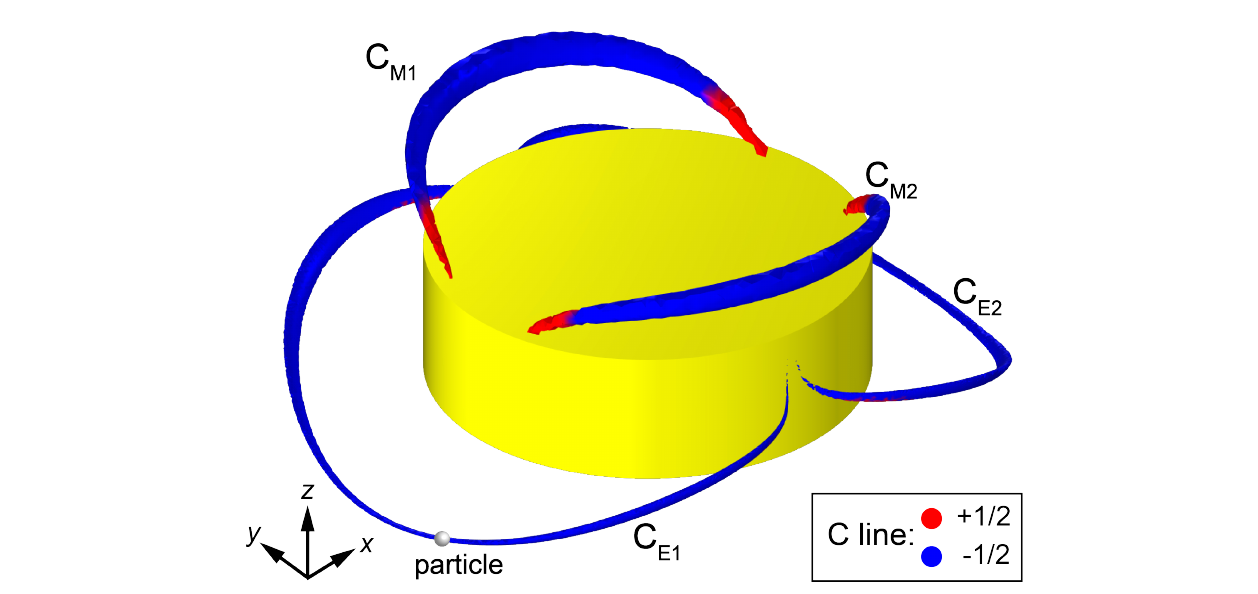}
\caption{The C lines in electric and magnetic near fields of a gold cylinder. The incident wave is linearly polarized along the $x$ axis and propagating along the $z$ axis. The electric C lines are labeled as $\mathrm{C}_{\mathrm{E} 1}$ and $\mathrm{C}_{\mathrm{E} 2}$. The magnetic C lines are labeled as $\mathrm{C}_{\mathrm{M} 1}$ and $\mathrm{C}_{\mathrm{M} 2}$.  The C lines with blue (red) color carry a topological index of $-1/2$ $(+1/2)$.}
\label{fig:1}
\end{figure}

We consider the C lines emerging in the electric and magnetic near fields of a gold cylinder, as shown in Fig. 1. The cylinder has radius $r=500$ nm and height $h=400$ nm. The material property of gold is characterized by a Drude model $\varepsilon_{\mathrm{Au}}=1-\omega_{\text{p}}^{2}/(\omega^{2}+i \omega \gamma)$, where $\omega_{\text{p}}=1.28\times10^{16}$ rad/s and $\gamma=7.10\times10^{13}$ rad/s \cite{olmon_optical_2012}. The cylinder is under the excitation of an incident plane wave propagating along $z$-axis and polarized in $x$ direction. The amplitude of the incident electric field is 1 V/m. We conduct full-wave numerical simulation of the system by using a finite-element commercial package COMSOL \cite{COMSOL}.  Figure 1 shows the C lines emerging in the near fields at the frequency of 100 THz, where the lines labeled as $\mathrm{C}_{\mathrm{E} 1}$ and $\mathrm{C}_{\mathrm{E} 2}$ denote a pair of C lines in the electric field, while the lines labeled as $\mathrm{C}_{\mathrm{M} 1}$ and $\mathrm{C}_{\mathrm{M} 2}$ denote a pair of C lines in the magnetic field. We notice that the C lines possess a symmetric configuration due to the symmetry of the system. Importantly, the electric C lines and the magnetic C lines do not entangle with each other, allowing us to explore their corresponding optical forces/torques independently.  The colors of the C lines indicate their topological indices: blue (red) color denotes an index of $I=-1 / 2 (I=+1/2)$, which are determined by employing method in Ref.\cite{berry_index_2004}. As seen, the electric C lines have the same index $I=-1/2$ and are connected to the cylindrical surface of the cylinder. In contrast, the magnetic C lines have a sign change of the topological index and are connected to the edges of the cylinder. 

\begin{figure}[t!]
\centering\includegraphics[width=1\linewidth]{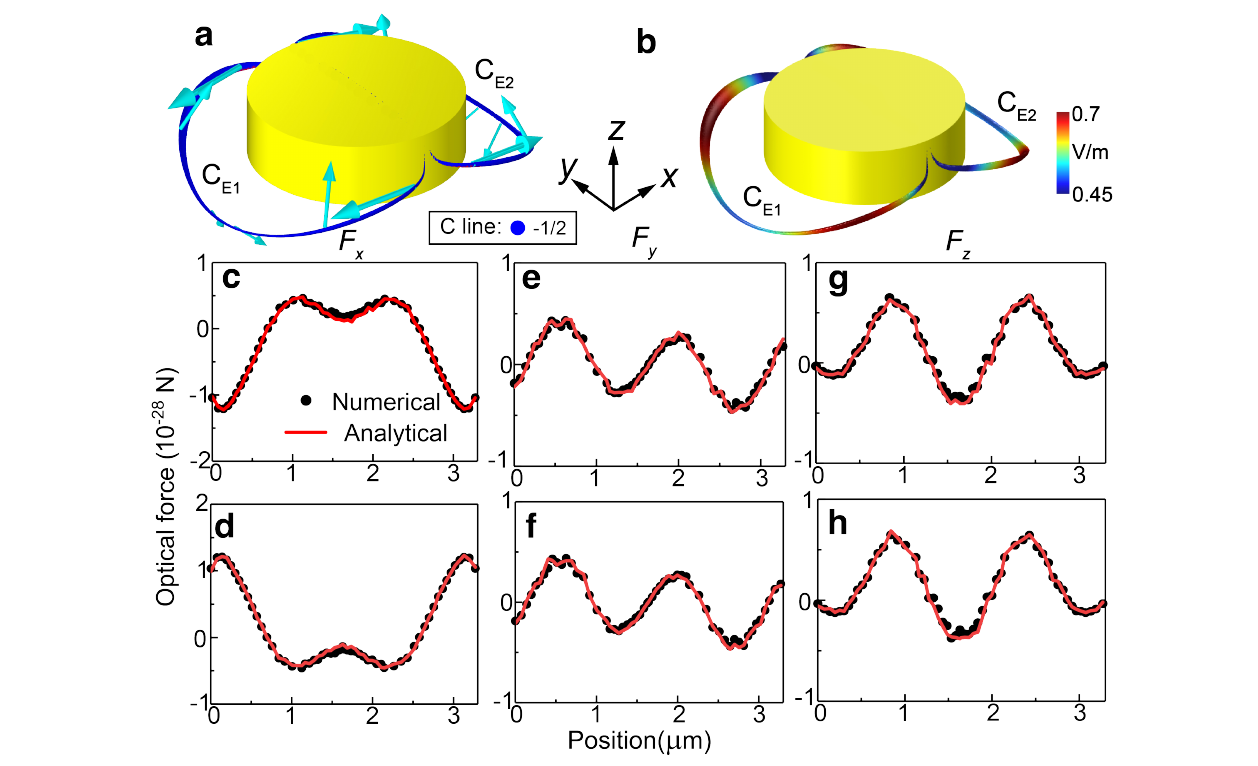}
\caption{(a) Optical force vectors (arrows) for a small dielectric particle ($\varepsilon_\mathrm{r}=4-0.1\text{i}$ and $\mu_\mathrm{r}=1$) locating on different positions on the electric C lines. (b) Amplitude of electric field on the C lines. Numerical (dots) and analytical (lines) results of the Cartesian components of the optical force induced by $\mathrm{C}_{\mathrm{E} 1}$ line (c, e, g) and $\mathrm{C}_{\mathrm{E} 2}$ line (d, f, h).}
\label{fig:2}
\end{figure}

\section{\label{Sec.3}Optical force and torque induced by the polarization singularities}
\subsection{\label{Sec. 3.1}Force and torque induced by electric C lines on a small dielectric sphere}
We consider a dielectric spherical particle with radius $a = 20$ nm locating on the electric C lines $\mathrm{C}_{\mathrm{E} 1}$ and $\mathrm{C}_{\mathrm{E} 2}$. The relative permittivity and relative permeability of the sphere is $\varepsilon_\mathrm{r}=4-0.1\text{i}$ and $\mu_\mathrm{r}=1$, respectively. The time-averaged optical force can be numerically calculated as \cite{jackson_classical_1999}:
    \begin{equation}
      \mathbf{F}=\oint \stackrel{\leftrightarrow}{\mathbf{T}} \cdot \hat{\mathbf{n}} dS,
        \label{eqn:1}
    \end{equation}
where $\stackrel{\leftrightarrow}{\mathbf{T}}=1 / 2 \operatorname{Re}[\varepsilon_{0} \mathbf{E}^{*} \mathbf{E}+\mu_{0} \mathbf{H}^{*} \mathbf{H}-1 / 2(\varepsilon_{0}|\mathbf{E}|^{2}+\mu_{0}|\mathbf{H}|^{2}) \stackrel{\leftrightarrow}{\mathbf{I}}]$ is the time-averaged Maxwell stress tensor, $\hat{\mathbf{n}}$ denotes the outward unit normal vector, $\stackrel{\leftrightarrow}{\mathbf{I}}$ is the unit tensor, and the integral is evaluated over a closed surface surrounding the sphere. We apply Eq.\ (\ref{eqn:1}) to evaluate the force on the sphere, and the results are shown in Fig.\ \ref{fig:2}. The arrows in Fig.\ \ref{fig:2}(a) show the force vectors for the sphere locating at different positions of the electric C lines. We notice that the force field undergoes dramatic variations along the C lines, indicating large field variations in the near field region. This is confirmed by the electric field amplitude shown in Fig.\ \ref{fig:2}(b). The quantitative numerical results of the component forces as a function of the position variable, defined as the accumulated distance from one end of the C line, are denoted by the black dots in Fig.\ \ref{fig:2}(c),(e),(g) for the sphere locating on $\mathrm{C}_{\mathrm{E} 1}$ and in Fig.\ \ref{fig:2}(d),(f),(h) for the sphere locating on $\mathrm{C}_{\mathrm{E} 2}$, respectively (the small fluctuations of the forces are due to the numerical errors in the spatial positions of the C lines). Assuming the origin of the coordinate system is at the center of the gold cylinder, the optical force in Fig.\ \ref{fig:2} possesses the following symmetries:
\begin{equation}
            F_{x}(-x)=-F_{x}(x), F_{y}(-x)=F_{y}(x), F_{z}(-x)=F_{z}(x), 
    \label{eqn:2}
\end{equation}
\begin{equation}
    F_{x}(-y)=F_{x}(y), F_{y}(-y)=-F_{y}(y), F_{z}(-y)=F_{z}(y).
            \label{eqn:3} 
\end{equation}
These symmetries are direct results of the symmetry of the cylinder and the excitation field, which are mirror symmetric with respect to the $yoz$-plane and $xoz$-plane. In this complex force field, the small particle will not be simply attracted to the metal cylinder as in many plasmonic trapping systems. The force can enable complicated manipulations of small particles in the near field of the cylinder.

To understand the physics behind the complex force field, we apply dipole approximation to the dielectric particle and analytically evaluate the optical force as \cite{chaumet_electromagnetic_2009,nieto-vesperinas_optical_2010,wang_lateral_2014}:
\begin{equation}
        \mathbf{F}=\frac{1}{2} \operatorname{Re}\left[\left(\nabla \mathbf{E}^{*}\right) \cdot \mathbf{p}+\left(\nabla \mathbf{H}^{*}\right) \cdot \mathbf{m}-\frac{c k^{4}}{6 \pi}\left(\mathbf{p} \times \mathbf{m}^{*}\right)\right],
    \label{eqn:4}
\end{equation}
where $\mathbf{E}$ and $\mathbf{H}$ are the electric and magnetic fields acting on the particle; $\mathbf{p}=\alpha_{\text{e}} \mathbf{E}$ and $\mathbf{m}=\alpha_{\text{m}} \mathbf{H}$ are the electric and magnetic dipoles induced in the particle, respectively. For a spherical particle, we have $\alpha_{\text{e}}=\frac{(i6\pi\varepsilon_0)}{k^3}  a_{1}$,  $\alpha_{\text{m}}=\frac{(i6\pi\mu_0)}{k^3}  b_{1}$, where $a_{1}$ and $b_{1}$ are the Mie coefficients \cite{bohren_absorption_2004}. The analytical results of the optical force are denoted by the solid red lines in Fig.\ \ref{fig:2}(c)-(h), which agree well with the full-wave numerical results. We find that the optical force is dominated by the first term in Eq.\ (\ref{eqn:4}) corresponding to the electric dipole force, which is orders of magnitude larger than the rest terms. This is because that the particle is made of dielectric material with $\mu_{\text{r}}=1$, and thus the magnetic polarizability is negligible in the deep-subwavelength regime. To further understand the origin of the force, we decompose the electric force into a gradient force and a scattering force as $\mathbf{F}=\mathbf{F}_\mathrm{g}+\mathbf{F}_\mathrm{s}=\nabla U+\sigma \mathbf{S}/c$, where $U=\frac{1}{4}\left(\operatorname{Re}\left[\alpha_\mathrm{e}\right]|\mathbf{E}|^{2}+\operatorname{Re}\left[\alpha_\mathrm{m}\right]|\mathbf{H}|^{2}\right)$ is the energy density, $\sigma=\sigma_\mathrm{e}+\sigma_\mathrm{m}=\frac{k_{0}}{\varepsilon_{0}} \operatorname{Im}\left[\alpha_{\mathrm{e}}\right]+\frac{\mathrm{k}_{0}}{\mu_{0}} \operatorname{Im}\left[\alpha_{\mathrm{m}}\right]$ is the total cross section, and $\mathbf{S}=\frac{1}{2} \operatorname{Re}\left[\mathbf{E} \times \mathbf{H}^{*}\right]$ is the time-averaged Poynting vector. We find that the electric force is dominated by the gradient force that is mainly attributed to the electric field gradient. This is confirmed by the numerical results of the $|\mathbf{E}|$ field shown in Fig.\ref{fig:2}(b). We notice that there are four points with local maximum of the electric field amplitude, in accordance with the directions of the force vectors in Fig. \ref{fig:2}(a).

\begin{figure}[t!]
\centering\includegraphics[width=1\linewidth]{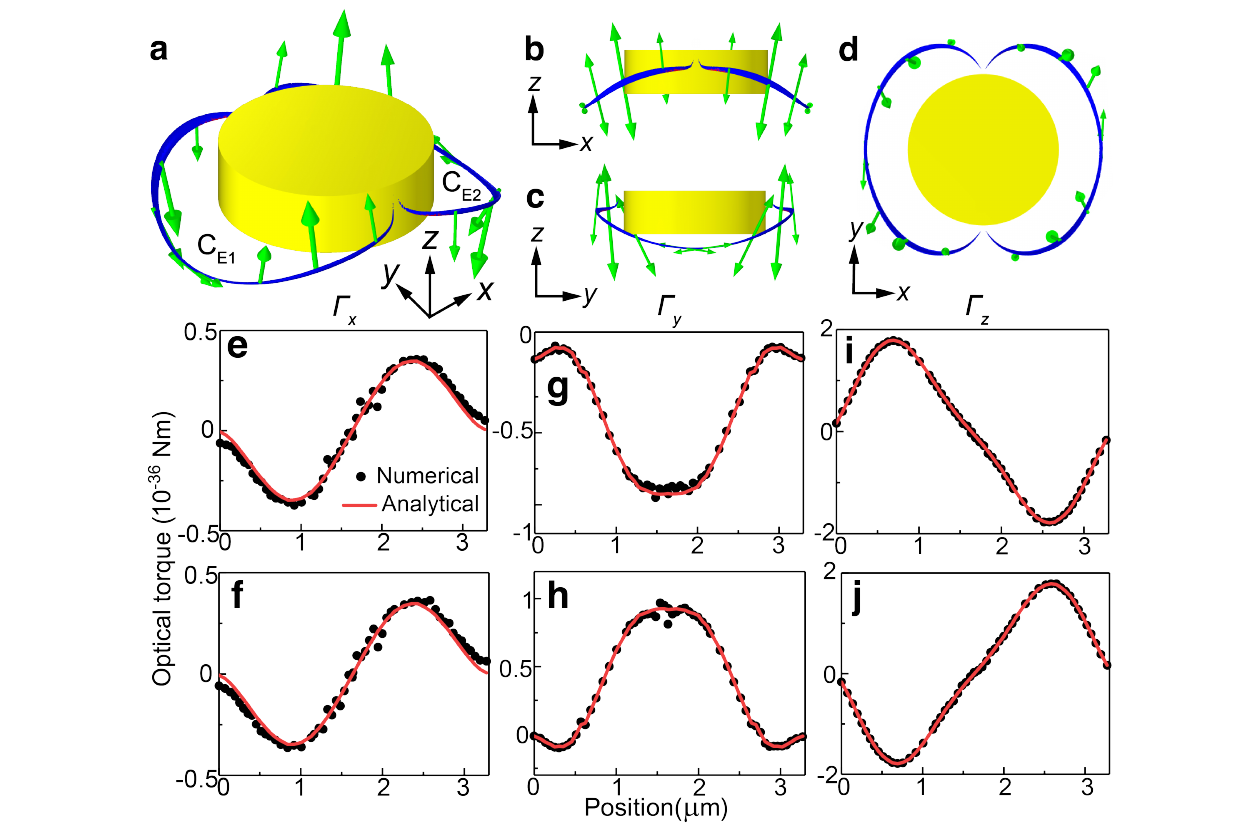}
\caption{(a-d) Optical torque vectors (denoted by the arrows) for a small dielectric particle ($\varepsilon_\mathrm{r}=4-0.1i$ and $\mu_\mathrm{r}=1$) locating on different positions on the electric C lines. Numerical (dots) and analytical (lines) results of the Cartesian components of the optical torque induced by $\mathrm{C}_{\mathrm{E} 1}$ line (e, g, i) and $\mathrm{C}_{\mathrm{E} 2}$ line (f, h, j).}
\label{fig:3}
\end{figure}

The optical torque on the small dielectric particle can be numerically calculated as
\begin{equation}
\mathbf{\Gamma}=\oint \mathbf{r} \times(\stackrel{\leftrightarrow}{\mathbf{T}} \cdot \hat{\mathbf{n}}) d S.
\label{eqn:5}
\end{equation}
The results for different locations of the particle on the electric C lines are shown in Fig.\ \ref{fig:3}. The green arrows in Fig.\ \ref{fig:3}(a)-(d) denote the torque vectors. Similar to the optical force, the optical torque also has large variations on the C lines. The quantitative numerical results of the Cartesian components of the torque are denoted by the black dots in Fig.\ \ref{fig:3}(e),(g),(i) for the sphere locating on $\mathrm{C}_{\mathrm{E} 1}$ and in Fig.\ \ref{fig:3}(f),(h),(j) for the sphere locating on $\mathrm{C}_{\mathrm{E} 2}$, respectively. We notice that the optical torque has the following symmetries:
\begin{equation}
\Gamma_{x}(-x)=\Gamma_{x}(x), \Gamma_{y}(-x)=-\Gamma_{y}(x), \Gamma_{z}(-x)=-\Gamma_{z}(x),
\label{eqn:6}
\end{equation}
\begin{equation}
\Gamma_{x}(-y)=-\Gamma_{x}(y), \Gamma_{y}(-y)=\Gamma_{y}(y), \Gamma_{z}(-y)=-\Gamma_{z}(y).
\label{eqn:7}
\end{equation}
Similar to the optical force, the above symmetries are attributed to the mirror symmetries of the system with respect to the $yoz$-plane and $xoz$-plane. Since torque is a pseudo-vector, under a mirror operation, the torque component perpendicular to the mirror plane remains unchanged, while the other two components undergo a change of sign. In the case of Eq.\ (\ref{eqn:6}) corresponding to the $yoz$ mirror plane,  $\Gamma_x$ is symmetric, while $\Gamma_y$ and $\Gamma_z$ are anti-symmetric. In the case of Eq.\ (\ref{eqn:7}) corresponding to the $xoz$ mirror plane, $\Gamma_y$ is symmetric, while $\Gamma_x$ and $\Gamma_z$ are anti-symmetric. These symmetry properties also agree with the symmetries of the spin density of the near fields.

Under the dipole approximation, the time-averaged optical torque on the dielectric particle can be analytically evaluated as 
\begin{equation}
\mathbf{\Gamma}=\frac{1}{2} \operatorname{Re}\left[\mathbf{p} \times \mathbf{E}^{*}+\mathbf{m} \times \mathbf{H}^{*}\right]=\frac{2 \omega}{\varepsilon_{0}} \operatorname{Im}\left[\alpha_{\mathrm{e}}\right] \mathbf{L}_{\mathrm{e}}+\frac{2 \omega}{\mu_{0}} \operatorname{Im}\left[\alpha_{\mathrm{m}}\right] \mathbf{L}_{\mathrm{m}},
\label{eqn:8}
\end{equation}
where $\mathbf{L}_{\mathrm{e}}=\frac{\varepsilon_{0}}{4 \omega} \operatorname{Im}\left[\mathbf{E}^{*} \times \mathbf{E}\right]$ and $\mathbf{L}_{\mathrm{m}}=\frac{\mu_{0}}{4 \omega} \operatorname{Im}\left[\mathbf{H}^{*} \times \mathbf{H}\right]$ are the time-averaged electric and magnetic spin densities, respectively. The analytical results given by Eq.\ (\ref{eqn:8}) are denoted by the solid red lines in Fig.\ \ref{fig:3}, which agree well with the numerical results. For the considered dielectric particle, the torque is mainly contributed by the electric spin density. The symmetries of the optical torque follow the symmetries of the electric spin density, which is also a pseudo-vector. The dramatic variations of the optical torque directly derive from the nonuniform electric spin density on the C lines. The complex distribution of the spin density gives rise to strong optical chirality in the near fields that drives the small particle to rotate in different directions depending on its position. These chiral near fields can also be employed for chiral discrimination \cite{jia_chiral_nodate}.

\begin{figure}[t!]
\centering\includegraphics[width=1\linewidth]{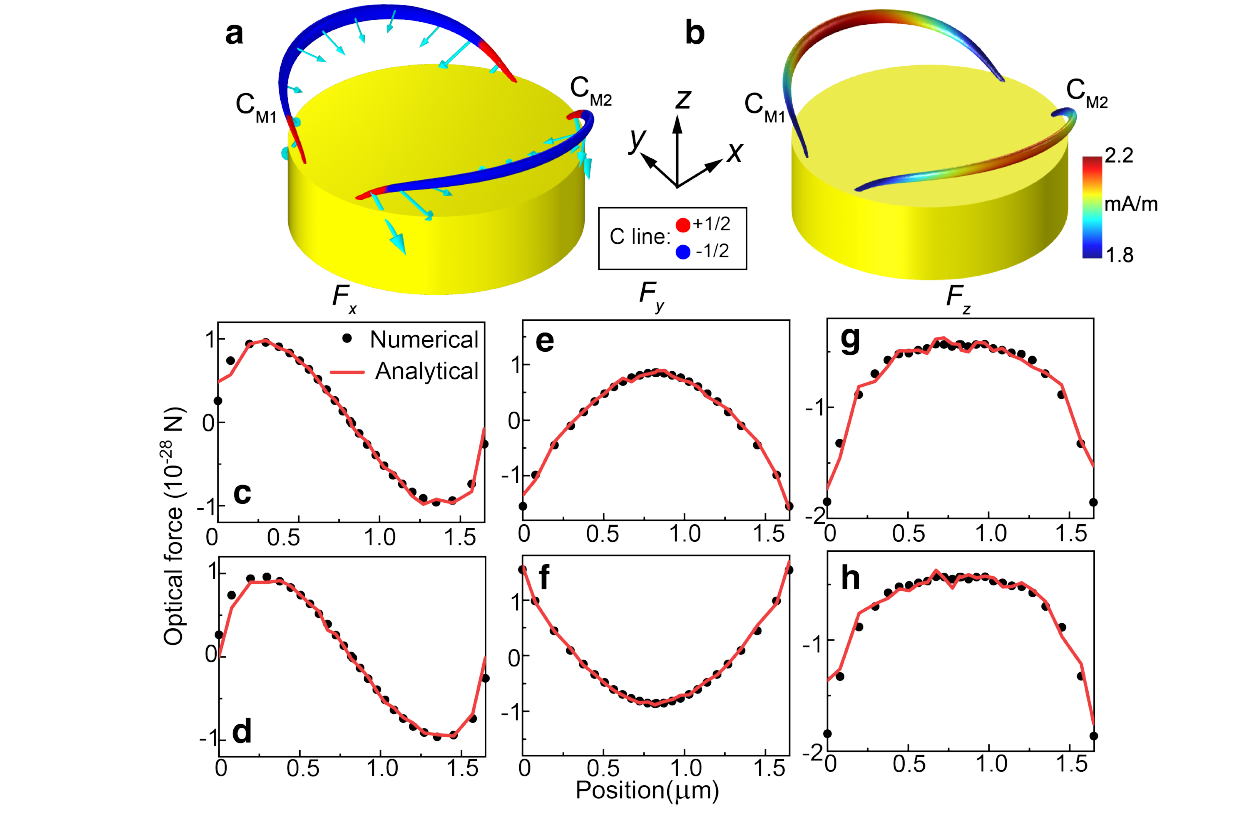}
\caption{(a) Optical force vectors (denoted by the arrows) for a small magnetic particle ($\varepsilon_\mathrm{r}=1$ and $\mu_\mathrm{r}=4-0.1i$) locating on different positions on the magnetic C lines. (b) Amplitude of the magnetic field on the C lines. Numerical (dots) and analytical (lines) results of the Cartesian components of the optical force induced by $\mathrm{C}_{\mathrm{M} 1}$ line (c, e, g) and $\mathrm{C}_{\mathrm{M} 2}$ line (d, f, h).}
\label{fig:4}
\end{figure}

\begin{figure}[t!]
\centering\includegraphics[width=1\linewidth]{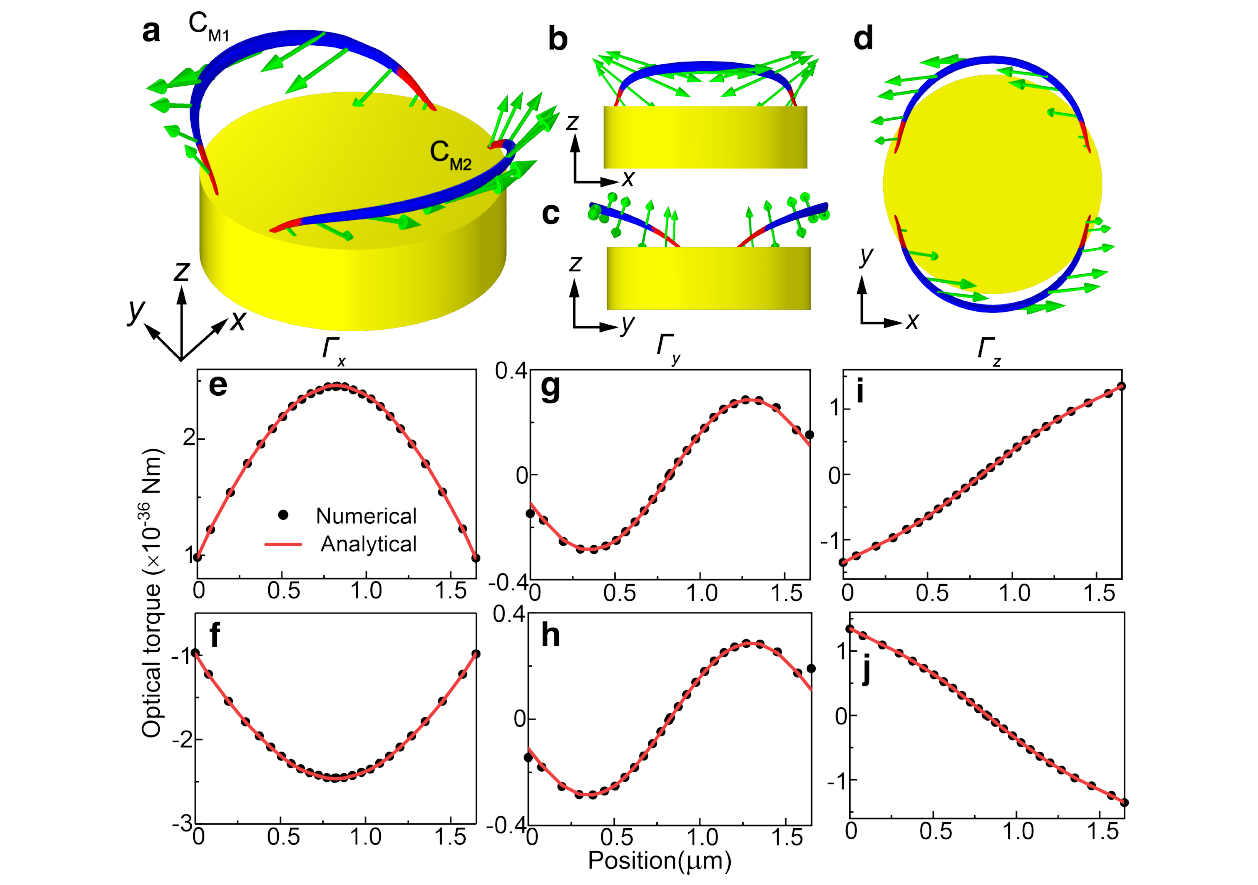}
\caption{(a-d) Optical torque vectors (denoted by the arrows) for a small magnetic particle ($\varepsilon_\mathrm{r}=1$ and $\mu_\mathrm{r}=4-0.1i$) locating on different positions on the magnetic C lines. Numerical (dots) and analytical (lines) results of the Cartesian components of the optical torque induced by $\mathrm{C}_{\mathrm{M} 1}$ line (e, g, i) and $\mathrm{C}_{\mathrm{M} 2}$ line (f, h, j)}
\label{fig:5}
\end{figure}

\subsection{\label{Sec. 3.2}Force and torque induced by magnetic C lines on a small magnetic sphere}
To understand the different properties of electric and magnetic C lines in optical manipulations, we also investigate the optical force and torque induced by the magnetic C lines $\mathrm{C}_{\mathrm{M} 1}$ and $\mathrm{C}_{\mathrm{M} 2}$. In this case, we consider a magnetic spherical particle with the same radius $a=20$ nm and relative permittivity  $\varepsilon_\text{r}=1$ and relative permeability $\mu_\text{r}=4-0.1\text{i}$. We apply the Maxwell stress tensor method to calculate the optical force, and the results are shown in Fig.\ \ref{fig:4}. The arrows in Fig.\ \ref{fig:4}(a) denote the optical force vectors for the sphere locating at different positions on the C lines. Compared to the force induced by electric C lines, the variation of the force due to the magnetic C lines is relatively smooth. This can be attributed to the fact that the magnetic near field is generated by the currents in the gold cylinder, while the electric near field is generated by the charges. The charges can give rise to diverging electric fields near the edges of the cylinder. Notably, the force is continuous at the transition points of the topological index $I=+1 \rightarrow-1$(corresponding to the points where the color changes from red to blue), indicating that the optical force is insensitive to the topological transition of the C lines.  The quantitative numerical results of the component forces are denoted by the black dots in Fig.\ \ref{fig:4}(c),(e),(g) for the sphere locating on $\mathrm{C}_{\mathrm{M} 1}$ and in Fig.\ \ref{fig:4}(d),(f),(h) for the sphere locating on $\mathrm{C}_{\mathrm{M} 2}$, respectively. As expected, the optical force manifests the same symmetries as in the case of Fig.\ \ref{fig:2}, i.e., the force is symmetric with respect to the $xoz$-plane and $yoz$-plane (assuming the origin is at the center of the cylinder). We also analytically calculate the optical force by using Eq.\ (\ref{eqn:4}), and the results are denoted by the solid red lines in Fig.\ \ref{fig:4}(c)-(h), which agree well with the numerical results. The force is dominated by the magnetic dipole force corresponding to the second term in Eq.\ (\ref{eqn:4}) due to the magnetic nature of the particle (i.e., $\mu_\text{r}=4-0.1\text{i}$), and it is mainly attributed to the gradient of the magnetic field. This is confirmed by the magnetic field $|\mathbf{H}|$ pattern shown in Fig.\ \ref{fig:4}(b), where the maximum values appear at the middle of the C lines and the field gradient along the C lines is small. This explains why the force vectors in Fig.\ \ref{fig:4}(a) are approximately perpendicular to the C lines.  

We apply Eq.\ (\ref{eqn:5}) to numerically calculate the optical torque acting on the small magnetic particle, and the results are shown in Fig.\ \ref{fig:5}. Figure\ \ref{fig:5}(a)-(d) shows the torque vectors in different views for the particle locating at different positions on the magnetic C lines. Compared to the torque induced by electric C lines, the optical torque in this case approximately points in the same direction on the same C line: $-x$ direction on the $\mathrm{C}_{\mathrm{M} 1}$ line and $+x$ direction on the $\mathrm{C}_{\mathrm{M} 2}$ line. Importantly, the direction of the torque is continuous at the transition points of the topological index $I=+1 \rightarrow-1$. This demonstrates that the optical torque is insensitive to the topological transition of the C lines. The quantitative numerical results are denoted by the black dots in Fig.\ \ref{fig:5}(e)-(j). The analytical results obtained with Eq.\ (\ref{eqn:8}) are denoted by the red lines, agreeing with the numerical results. We find that the torque is mainly attributed to the magnetic spin density $\mathbf{L}_\text{m}$, and it has the same symmetry properties as the torqued induced by the electric spin density $\mathbf{L}_\text{e}$, as expressed in Eqs.\ (\ref{eqn:6})-(\ref{eqn:7}).

\section{\label{Sec.4}Conclusion}
We discuss the optical force and torque on small dielectric/magnetic particles induced by polarization singularities (i.e., C lines) emerged in electric and magnetic near fields of a metal cylinder. It is demonstrated that the C lines can give rise to interesting configurations of the force and torque fields. The force and torque induced by electric C lines have dramatic variations enabling rich patterns of manipulations. In contrast, the force and torque fields attributed to the magnetic C lines are relatively smooth. Thus, two types of C lines can be selectively employed to manipulate small particles in different scenarios. By applying dipole approximations, we show that the forces are dominated by the gradient force of the dipoles induced in the particles, and the torque is directionally proportional to the electric/magnetic spin densities. Our study indicates that even a simple metal cylinder can give rise to rich degrees of freedom for near-field optical manipulations. The results bridge the field of singular optics with optomechanics and can find applications in chiral light-matter interactions, on-chip optical manipulations, and optical sensing.

\begin{backmatter}
\bmsection{Funding}
Research Grants Council of the Hong Kong Special Administration Region, China (CityU 11306019); National Natural Science Foundation of China (11904306).
\bmsection{Disclosures}
The authors declare no conflicts of interest.
\bmsection{Data availability}
Data underlying the results presented in this paper are not publicly available at this time but may be obtained from the authors upon reasonable request.
\end{backmatter}

%%%%%%%%%%%%%%%%%%%%%%% References %%%%%%%%%%%%%%%%%%%%%%%%%

\bibliography{sample}

\end{document}